\documentclass[prd,twocolumn,floatfix]{revtex4-2}

\newcommand{\omitfigs}{0}

\usepackage{amsmath}  
\usepackage{amsfonts}  
\usepackage{graphicx}  
\usepackage{epstopdf}  
\usepackage{amssymb}

\usepackage{epstopdf}

\usepackage{pgfplots}
\usetikzlibrary{pgfplots.groupplots}

\newcommand{\mytikz}[2]
{\ifodd \omitfigs
\else
\begin{tikzpicture}[scale=#1] \input{#2} \end{tikzpicture}
\fi
}

\newcommand{\mytikzoct}[2]
{ \centerline {\resizebox{!}{#1\textwidth}{ \input{#2} }  } }

\newcommand{\be}{\begin{eqnarray}}
\newcommand{\ee}{\end{eqnarray}}

\newcommand{\brho}{ b}
\newcommand{\vx}{{\bf x}}

\newcommand{\dt}{{\rm d}t}
\newcommand{\kB}{k_{\rm B}}

\newcommand{\dby}[2]{ \frac{{\rm d} #1}{{\rm d} #2}}

\newcommand{\nB}{ n_{\scalebox{.8}{$\scriptstyle \rm B$}} }

\begin{document}
\def \d {{\rm d}}

\ifodd 0
\fi

\title{Gravitational bremsstrahlung in plasmas and clusters}

\author{Andrew M. Steane}
\email{a.steane@physics.ox.ac.uk} 
\address{Department of Atomic and Laser Physics, Clarendon Laboratory, Parks Road, Oxford OX1 3PU, England.}

\date{\today}

\begin{abstract}
We study the gravitational bremsstrahlung owing to collisions 
mediated by a $1/r$ potential. We combine classical
and first order Born approximation results in order to construct an 
approximate gravitational `Gaunt factor' for the total emitted
energy. We also obtain the cross-section with an angular momentum
cut-off, and hence the cross-section for emission via close hyperbolic
encounters in a gravitating cluster. 
These effects are the dominant source of very high frequency 
gravitational noise in the solar system. The total 
gravitational wave power of the Sun is $76\pm 20\,$MW. 
\end{abstract}







\maketitle

This paper reviews and extends the study of gravitational bremsstrahlung during collisions in a $1/r$ potential. 
In practice this is Coulomb collisions and 
gravitational `collisions' (i.e. hyperbolic encounters) 
where the potential is well-approximated as $1/r$.  
Such processes take place in plasmas such as stellar interiors, and
in gravitating clusters such as those of black holes believed to be
present in many galactic nuclei, or in the early universe. 
However the motivation to study these processes is mainly their innate interest.
They involve a combination of quantum theory and dynamic gravitation. 
For Coulomb collisions in the Sun
the resulting gravitational wave amplitude is small and undetectable on Earth using
any technology liable to be realised in the near future, but in principle
it contributes to the limits on coherence of matter-wave interferometry 
owing to gravitational field noise.\cite{Steane2017,Cutler2002,Aggarwal2021,00Maggiore}

Introductory material is set out in the first two sections below.
Section \ref{s.history} surveys previous work on
gravitational wave (GW) emission during collisions in a $1/r$ potential
at low (non-relativistic) speeds.
Section~\ref{s.not} introduces notation and methods.
Section~\ref{s.total} obtains the total cross-section for the GW energy
emission after integrating over impact parameter. This consists in
reporting existing work treating classical and quantum (first order Born approximation) limits, and providing approximate 
formulae for the intermediate regime.
Section~\ref{s.single} considers emission during a
single hyperbolic encounter.
Section \ref{s.cutoff} presents the cross-section obtained if one imposes
a cut-off on the angular momentum. This is useful for 
the case of attractive forces, where it makes sense to
separate the collisions into those leading to capture and those where
the bodies escape to infinity.
Section~\ref{s.cluster} obtains the GW energy emission cross-section for close hyperbolic encounters
in a gravitating cluster.
Section~\ref{s.sun} estimates the
total GW power of the Sun. Section \ref{s.conc} concludes.


\section{Historical survey} \label{s.history}

Early work on graviton emission during scattering of fundamental particles
was carried out by Ivanenko and Sokolov (1947, 1952). \cite{Ivanenko1947,Ivanenko1952}.
In 1965 Weinberg calculated gravitational bremsstrahlung during Coulomb
collisions using quantum field theory, in the limit where the gravitons are `soft',
meaning they have negligible impact on the energy-momentum in 
lines of Feynman diagrams on or near the mass shell.\cite{Weinberg1965} The following year Carmeli confirmed this and
also provided a classical calculation, for a repulsive potential,
finding the total emitted energy after integration over impact parameters.\cite{Carmeli1967}
His clever method of calculation did not require an expression for the emitted
energy in each hyperbolic encounter. Boccaletti (1972) extended this method
to the Yukawa potential, and estimated emission from
neutron stars.\cite{Boccaletti1972} 
Meanwhile Barker {\em et al.} 1969 gave the Born approximation
calculation for graviton emission during collisions in a $1/r$ potential, among
other results.\cite{Barker1969}.
Emission from binary stars on Keplerian orbits
had also been calculated, pioneered by Peters and Matthews (1963).
\cite{Peters1963,Peters1964}.

The above all concern low velocities and Euclidean geometry. Pioneering
calculations for the case of a Schwarzschild-Droste metric and arbitrary
velocity were provided by Peters (1970).\cite{Peters1970} Since then there
has been a very large body of work devoted to post-Newtonian corrections
to orbits and scattering, with impressive recent progress, 
see for example the remarkable \cite{Herrmann2021,Barack2022} and
references therein. In the present survey we will not
pursue the high-velocity or non-Newtonian cases. 
We restrict to $v \ll c$ and $r \gg R_{\rm S}$ 
for participating massive objects (with $R_{\rm S}$ the Schwarzschild radius)
and the quadrupole approximation applies.

Gal'Tsov and Grats (1974) carried out Born approximation calculations, giving
some further information not included in Barker {\em et al.}.\cite{Galtsov1974} They subsequently
(1983) extended their study towards a more complete kinetic theory of a plasma
such as that of the Sun.\cite{Galtsov1983}

The first person to have correctly reported the total GW energy emitted during
a hyperbolic encounter in a $1/r$ potential, according to 
classical (not quantum) physics,
appears to be Turner (1977), correcting a minor error
in a previous calculation by Hansen.\cite{Turner1977,Hansen1972} 
This work was duly noted in a comprehensive
review by Kovacs and Thorne in 1978, who comment:
``Such computations are straightforward and simple," but in view of the fact that
errors exist in the literature (we will point out some more in the next paragraphs) 
such computations are clearly not straightforward for ordinary mortals.\cite{Kovacs1978}

Dehnen and Ghaboussi 1985 treated a general central potential and report 
some useful results for that case.\cite{Dehnen1985,Dehnen1985B}
They apply their methods to the $1/r$ potential
as an example and obtain the
total scattered energy. Their formula agrees with that of Turner.
They did not cite Turner, presumably an indication that they were not aware
of his work. 
(Different authors report the formula in terms of different parameters so the agreement is not self-evident; we shall display both versions in section \ref{s.single}.)

Further reviews of astrophysical sources of gravitational waves are provided by
\cite{Papini1977,Cutler2002,Aggarwal2021}. 
Whereas Papini and Valluri
discuss bremsstrahlung inside stars along with other processes, Cutler and Thorne
do not because their review is focussed on signals that may be detectable now or in the near future. 

Recently a further case has gained interest: the emission from clusters of
black holes which may have been produced in the early universe or in the centres
of galaxies. 
\cite{Capozziello2008,OLeary2009,Vittori2012,GarciaBellido2018,GBellido2022}
The emission is partly from masses
in bound orbits, and partly from a background of close hyperbolic encounters.
In this work we are concerned with the latter, because it has received less
attention in the literature and because it can, in principle, 
dominate, depending on the parameters of any given cluster.
Capozziello {\em et al.} (2008) calculated the power and total
emitted energy per encounter in the case $r \gg R_{\rm S}$ where the
gravitational potential is Newtonian to good approximation.
Their results reproduce those of Turner and of
Dehnen and Ghaboussi though they cite neither; they cite the
review by Kovacs and Thorne which includes Turner but they do not make the
comparison. De Vittori {\em et al.} (2012) follow the method of Capozziello
explicitly but their eqn (6) has a sign error in the last term and their
eqn (8) has the total power too large by a factor 4. 
Garc\'{i}a-Bellido and Nesseris, and also
Gr\"{o}bner {\em at al.},
point out further mistakes. In view of these discrepancies a new calculation
may be useful and we provide one. 

The spectrum of the emitted radiation was treated by various authors,
with noteworthy contributions from Turner, 
O'Leary {\em et al.}, De Vittori {\em et al.},
Garc\'{i}a-Bellido and Nesseris
and Gr\"{o}bner {\em at al.}. (Gr\"{o}bner {\em et al.}'s opening statement 
that De Vittori {\em et al.} constitutes
`the first calculation of the frequency spectrum' 
understates the contribution of Turner who gave explicit formulae for
the cases of eccentricity
$e=1$ and $e \gg 1$ and much of the analysis for general
$e$; subsequent authors completed  the
Fourier integrals for all $e$). Some mistakes in \cite{Vittori2012} are corrected in \cite{GarciaBellido2018,Grobner2020}.

The existing studies
for electrical plasmas and those for gravitating clusters
appear to be unaware of one another although they are often calculating
the same things (i.e. emission during scattering in a $1/r$ potential).
The present work does the following:
(i) bring together the two communities just outlined; 
(ii) present the work of Galt'sov and Grats afresh; 
(iii) estimate the case, intermediate between classical and quantum, which
is not amenable to classical nor Born approximations, obtaining
an approximate `Gaunt factor' for the total emitted power;
(iv) obtain an emission cross section by using an angular momentum cut-off;
(v) show how the above can be applied to calculate the emission from
gravitating clusters and from a stellar plasma.



\section{Notation and general approach} \label{s.not}

For two colliding partners of masses $m_1$, $m_2$ we define
the total mass $M = m_1 + m_2$ and the reduced mass $\mu = m_1 m_2/M$.
We shall also use the unadorned $m$ (with no subscript) 
for reduced mass; thus $m \equiv \mu$.
A given binary collision is described in the COM frame, such that it
consists in a particle of mass $\mu$ moving in a fixed central
potential of the form either $V(r) = Z_1 Z_2 e^2 / r$ or $V(r) =
- G m_1 m_2 / r$. It is only necessary to treat one of these two cases since
the other is then given by making the replacement $Z_1 Z_2 e^2 \leftrightarrow -G m_1 m_2$. In the following we mostly present the
former (Coulomb scattering) since it includes both attractive
and repulsive collisions, and also preserves in the notation the
distinction between the potential and the role 
of $G$ in the GW emission process. 
For a slightly more succinct notation we define
$e_1 e_2 \equiv Z_1 Z_2 e^2$. 
We adopt electromagnetic units such that the Coulomb force between
electrons is $e^2/r^2$ and the fine structure constant 
is $\alpha = e^2 /\hbar c$.


For a collision with the masses initially far apart, $v_0$ is the
initial velocity and $b$ is the impact parameter. The collision energy
is $E = (1/2) \mu v_0^2$ and angular momentum $L = \mu b v_0$. 

If a flux $n_2 v$ is incident on a single collision centre, then the
rate of collisions is $n_2 v \sigma$ where $\sigma$ is the cross section
(this defines $\sigma$). 
If there is a density $n_1$ of collision centres, then the collision
rate per unit volume is $n_1 n_2 v \sigma$ if the particle types 1 and 2
are distinct, and it is $(1/2) n_1^2 v \sigma$ if the particle types are
not distinct. In this paper we shall write $n_1 n_2 v \sigma$ and expect
the reader to understand that in the case of identical particles the
factor half must be introduced.

Our discussion is entirely non-relativistic. This
is a good approximation for the core of the 
Sun, where the Lorentz factor 
$\gamma \simeq 1.004$ for electrons.

Gravitational bremsstrahlung has some features in
common with electromagnetic bremsstrahlung, which has been studied
extensively. For the latter,
the emitted
power per photon solid angle and frequency range 
is written as a product of an approximate classical
expression and a factor $g_{\rm ff}$ called the ``free-free {\em Gaunt factor}"
which incorporates quantum and other corrections. Complicated
expressions exist for $g_{\rm ff}$ but for many purposes it is useful
to have a simpler formula of reasonable accuracy. For the electromagnetic
case this has recently been provided by Weinberg \cite{Weinberg2019}. 

For an approximate classical calculation, one way to proceed is to 
integrate the emitted power 
at each moment for a particle
moving on the trajectory it would follow if
no radiation were emitted. For GW emission this approximation
holds very well for particle collisions and we shall adopt it.

Whether in the electromagnetic or GW case, there are two
significant energy scales in the collision dynamics: 
the kinetic energy and the 
potential energy at a distance of order a de-Broglie wavelength.
The former is $(1/2)m v^2$ where $v$ can be taken as the speed
at infinity for a repulsive potential, or as the speed at
the distance of closest approach for an attractive potential. 
For low angular momentum the
speed and acceleration have very different behaviours for
attractive and repulsive cases, leading to different
GW emission even though the differential cross section 
of the collision may be
independent of the sign of the potential. 

For Coulomb collisions between particles of charges $Z_1e,\,Z_2e$
we define the dimensionless parameter $n_{\rm B}$ called the
{\em Born parameter} by Galt'sov and Grats (and called $\xi$ by Weinberg  \cite{Weinberg2019}):
\be
\nB \equiv \frac{ |Z_1 Z_2 e^2| }{\hbar v} = |Z_1Z_2| \alpha \frac{c}{v}
\ee
The Born parameter can be read as a statement either about energy or about angular momentum. 
It is the ratio of the Coulomb energy at $2 \lambdabar_{\rm dB}$ to
the collision energy. It is also approximately equal to the angular momentum in units of $\hbar$. For a repulsive potential the
distance at closest approach is $2 \nB \lambdabar_{\rm dB}$
according to classical mechanics. 
The case $\nB \lesssim 1$ is the quantum limit; the
Born approximation for the scattering holds when $\nB \ll 1$. 
The case $\nB \gg 1$ is
the classical limit. Thus low temperatures give classical trajectories.
The ground state of hydrogen has $\nB \approx 1$. 

A further relevant energy is that of the emitted photons or gravitons,
$h \nu$. We say the photons or gravitons are `soft' when $h\nu \ll
(1/2) m v^2$ and `hard' otherwise. The maximum possible emitted photon
or graviton energy is equal to the entire kinetic energy 
$(1/2) m v^2$. More generally if a single
photon or graviton is emitted then the initial and final
momenta of the scattered particle 
(e.g. electron) in the COM frame are related by
\be
\frac{p_i^2}{2 m} - \frac{p_f^2}{2 m} = h \nu.
\ee
The collision process itself has a timescale
$\tau \approx r_0 / v$ where $r_0$ is the distance of closest approach.
Classical mechanics predicts that the emitted spectral power extends up
to the angular frequency range near $1/\tau$, but quantum mechanics 
gives a hard cut-off at $\omega = (1/2)mv^2/\hbar$. The question arises, then, whether
the classically `preferred' frequency is available. 
The condition that $1 /\tau$ is less than
the cut-off is $2 \hbar < m v r_0$, i.e. $\nB  > 1$.

A further consideration in the gravitational case is whether or not the $1/r$ 
potential is a good approximation. This can be expressed by the condition $r \gg 
R_{\rm S}$ already mentioned. We will assume this condition holds, and we show at
the end (eqns (\ref{eminus})--(\ref{emep})) that for the orbits under study 
(namely, non-captured hyperbolic orbits) this can be subsumed under the condition
$v_{\rm max} \ll c$. The reason why this is sufficient, without a further 
constraint on the impact parameter, is that a constraint is placed implicitly 
by restricting attention to orbits which do not undergo radiative capture. In 
other words, slow encounters can only avoid capture by remaining at large 
distances. The question, whether the methods offer a good approximation in any given case, then becomes a matter of practical astronomy concerning the velocity- and position-distributions in any given cluster.

\subsection{Methods of calculation}  \label{s.gravrad}

In the compact source approximation in linearized gravity, the luminosity 
(i.e. the emitted power) of a source is given by
\be
L_{\rm GW} = \frac{G}{5 c^5} \left< \dddot{Q}_{ij} \dddot{Q}^{ij} \right>
\label{LGW}
\ee
where 
\be
Q^{ij}  = \frac{1}{c^2} \int T^{00} (x^i x^j - \frac{1}{3} \delta_{ij} x^k x_k) \, \d^3 \vx
\ee
is the quadrupole moment of the mass distribution and
the angle bracket indicates an average over a small region of spacetime.

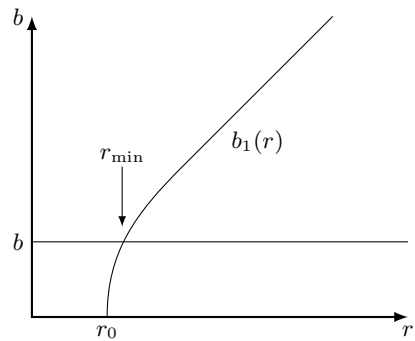
\begin{figure}
	\centering		
\begin{tikzpicture}
\tikzset{>=latex}
  \draw[thick, <->] (0,4) to (0, 0) to (5,0);   
  \draw[] (0,1) to (5,1);
  \draw[] (1, 0) to [out=90, in=225] (2,2) to [in = 225] (4,4);
  \node[left] at (0, 1) {$\brho$};
  \node[left] at (0, 4) {$\brho$};
  \node[below] at (1, 0) {$r_0$};
  \node[below] at (5, 0) {$r$};
  \node[below] at (3,2.6) {$\brho_{1}(r)$};

  \draw[->] (1.2,2) to (1.2,1.2);
  \node[above] at (1.2, 1.95) {$r_{\rm min}$};

\end{tikzpicture}
\caption{The region of integration of (\ref{intclass}) and 
(\ref{intLclass}). $\brho$ is the impact parameter, $r$ is the distance
from the origin in the COM frame. 
At any given impact parameter $\brho$, the trajectory
does not reach values of $r$ below $r_{\rm min}$ 
and therefore at any given $r$ it does not reach values of $\brho$
above $\brho_{1}$.}
\label{f.impact}
\end{figure}

For given collision partners, a collision is parametrised by
two quantities: the initial velocity $v_0$ and the impact parameter $\brho  $.
We can express the total power generated in some small volume $V$ of
a plasma, as a result of collisions between particles of types 1 and 2, as
\be
P = V n_1 n_2 \left< v_0 \Sigma \right>          \label{PSigma}
\ee
where $n_1$ and $n_2$ are number densities of two species
($n_1 n_2$ should be replaced by $(1/2)n_1^2$ if the species
are identical, as already remarked) and $\Sigma$ is a cross section
(to be calculated) with the physical dimensions of energy times area.

We shall obtain $\Sigma$ by calculating the total GW
energy emitted during a single collision, integrated over impact
parameter $\brho  $. We adopt and
compare four methods of calculation, as follows.
\begin{enumerate}
\item {\bf Purely classical}.
We calculate that trajectory in the COM frame. The total
emitted energy is $\int L_{\rm GW} \dt$ per collision,
with $\dddot{Q}_{ij}$ obtained from the trajectory.
The GW emission cross section is
\be
\Sigma  = \int_{-\infty}^\infty \dt
\int_0^\infty 2 \pi \brho  \, \d\brho   \, L_{\rm GW}
\ee
By exploiting the
symmetry of the inward and outward motion this can be written
\be
\Sigma = 2 \, \int_{r_0}^\infty \frac{\d r}{|\dot{r}|} 
\int_0^{\brho  _{\rm max}} 2 \pi \brho  \, \d\brho   \, L_{\rm GW}
\label{intclass}
\ee
where $r_0$ is the smallest distance of closest approach and
$\brho  _{\rm max}$ is the largest impact parameter whose associated
trajectory can reach a given $r$; see Fig. \ref{f.impact} for
an elucidation of this.

\item {\bf Born approximation}.
For a quantum treatment in the Born approximation
we shall present results obtained by Barker {\em et al.} and by
Gal'tsov and Grats (GG).\cite{Barker1969,Galtsov1974}

\item {\bf Soft photon theorem}.
Weinberg has obtained a very general expression for the emission
of soft massless particles in any collision. In the non-relativistic
limit his `soft photon theorem' applied to gravitons
yields an expression 
for the power in the radiated spectrum up to frequency
$\Lambda/\hbar$:
\be
P_{< \Lambda} \simeq V \frac{8 G}{5\pi c^5} m^2 v^5 n_1 n_2 \frac{\Lambda}{\hbar}
\int \dby{\sigma}{\Omega} \sin^2 \theta \, \d\Omega
\label{soft}
\ee
where $\Lambda$ is an energy cut-off which has to be taken low
enough so that it is small compared to relevant kinetic energies
in the problem, and $\d\sigma/\d\Omega$ is the differential cross
section for the collision in the absence of radiant emission. 
The term `soft' here means the graviton's energy-momentum is small 
compared to that of the particle emitting it.

Weinberg's formula does not give the whole emitted power,
only the part owing to soft gravitons, and only that part up the
frequency cut-off $\Lambda/\hbar$. Therefore
we should not expect it to match calculations
of the whole power. Nonetheless it offers a useful consistency 
check.
Expressed as a cross-section we have
\be
\Sigma_{< \Lambda} \simeq  \frac{8 G}{5\pi c^5} m^2 v^4  \frac{\Lambda}{\hbar}
\int \dby{\sigma}{\Omega} \sin^2 \theta \, \d\Omega \, .
\label{Sigsoft}
\ee

\ifodd 0
The soft photon (or graviton) theorem concerns gravitons attached
to external legs of a Feynman diagram and which do not significantly
change the momentum. Here `external' means lines for which the 4-momentum is near the mass shell.
This is a useful method for repulsive potentials
where the particles have their highest momentum in the initial and final states. For an attractive potential, however, Eqn (\ref{soft}) is
less useful in the classical limit, as we shall see.

The above formula implies that the emitted spectrum is uniform
over frequency, and this is indeed the prediction for soft gravitons
at low particle energies. For general particle energies the theorem
gives the low-frequency part of the spectrum as
$\omega^B$ where $B$ is a function of velocity which is of order
$G \ddot{Q}_{ij}^2/\hbar c^5$; this
is very small ($< 10^{-38}$) for collisions of fundamental particles.
\fi

\item {\bf Modified classical}.
With a view to gaining intuition about the quantum limit, and
to obtain formulae which are approximately valid for any initial
velocity, we explore the effect of modifying the classical formula
(\ref{intclass}). This is not a modification to the equation of motion;
it is merely a rough method to gain reasonable insight and approximate
formulae. 
The idea is that the quantum behaviour can be modelled roughly
by using a classical mass distribution with mass density equal to 
$m \left| \psi \right|^2$ where $\psi$ is a wavefunction in position space, and
we suppose this distribution has a peaked (e.g. Gaussian) form with a standard deviation
to be discovered and a mean which follows the classical trajectory. We then suppose
that, to sufficient approximation, the result of such a model
can be estimated by some simple adjustment to the integrand in (\ref{intclass}).

One idea, for example, is to replace $r$ in the integrand
of (\ref{intclass}) with some simple function such as $(r^2 + \Delta^2)^{1/2}$
where $\Delta$ is a parameter to be set so as to reproduce the known
behaviour in the limits of small and large Born parameter. One would
expect this $\Delta$ to be of the order of the de Broglie wavelength.
This was explored, and so were other possibilities. In particular,
one might leave the integrand unchanged and simply adjust the lower limit of
the integral. It was found that this
gives a good approximation. This is presented in
sections \ref{s.semi},~\ref{s.cutoff}.


\end{enumerate}

\section{Total emission cross section}
\label{s.total}

\subsection{Order-of-magnitude estimate}

In order to get some general insight into the results to be discussed,
we first present a simple order-of-magnitude estimate of GW radiation
during repulsive Coulomb collisions.

From (\ref{LGW}) we have
\be
L_{\rm GW} \approx \frac{G}{5 c^5} \left(\frac{\overline{M x^2}}{\tau^3} \right)^2
\approx \frac{4 G}{5 c^5} \left(\frac{E_Q}{\tau} \right)^2
\ee
where $\tau$ is the timescale and $E_Q$ is the part of the kinetic energy associated
with non-spherical (i.e. quadrupolar) movements. 
The timescale of the changing quadrupole moment is $\tau \simeq  r_0 / v$
where $r_0$ is a characteristic distance scale for a collision at energy $E$ and
$v$ is the relative speed of the colliding partners.
We take $r_0$ equal to the distance of closest approach in a head-on collision, 
\be
r_0 = e_1 e_2 / E  =  \frac{2 e_1 e_2}{\mu v_0^2}.     \label{r0}
\ee

The duration of each collision is about $2\tau$ so the emitted energy
per collision is $(8G/5c^5) E^2 / \tau$. 
Multiplying this by the collision
rate $n_2 \sigma v$ and the number density $n_1$, and using $\sigma = 4\pi r_0^2$, we obtain the power per unit volume of the gravitational
wave production:
\be
\frac{P}{V} \approx n_1 n_2 e_1 e_2
\frac{64 \pi G}{5 c^5} \frac{E^2}{\mu}.
\label{PeeEst}
\ee

Eqn (\ref{PeeEst}) is compared with the result of a precise
calculation in the next section. We there find
that it captures correctly the scaling with parameters 
of the classical result for a repulsive potential, and gets the
numerical factor about right.

\subsection{Classical treatment}

We treat the two-body dynamics as a single-body motion of
a particle of mass $\mu$ moving in a static
potential centred on the origin. Let $D_{ij} \equiv 3 Q_{ij}$,
then $D_{ik} = \mu (3 x_i x_k - x^j x_j \delta_{ik})$ and
\be
\ddot{D}_{ik} = 6 \mu v_i v_k - 6 \dby{V}{r} \frac{1}{r} x_i x_k
- 2\left[ \mu v_j v^j -  \dby{V}{r} \frac{1}{r} x^j x_j \right] \delta_{ik}.
\nonumber
\ee
The calculation of $\dddot{D}_{ik} \dddot{D}^{ik}$ is straightforward
and the result is given by Boccaletti. \cite{Boccaletti1972} 
For Coulomb collisions one finds
\be
L_{\rm GW} = \frac{8G}{15 c^5} \frac{(e_1 e_2)^2}{r^4}\left( v^2 + 11 v_\perp^2 \right)                   \label{Powervv}
\ee
where $v_\perp^2 = v^2 - \dot{r}^2$ and in this expression $v,\,v_\perp$
and $r$ are all functions of time. 

The case of gravitational scattering can be treated by the replacement
$e_1 e_2 \rightarrow -G m_1 m_2$. 

The potential is 
\be
V(r) = e_1 e_2 / r \, ,
\ee
which may be positive or negative, depending on the signs of the charges.
In the case of a repulsive force (potential hill), $r_0$ 
is a positive number equal to the minimum distance attained
in a head-on collision. In the case of an attractive force (potential well)
$r_0$ has no such interpretation but we retain the formula 
(\ref{r0}) as a definition, and then $r_0 < 0$. 

From conservation of energy and angular momentum we have
\be
v^2 &=& v_0^2(1 - r_0/r) , \qquad
v_\perp = v_0 \brho   / r            \label{conserve}
\ee
where $v_0$ is the initial velocity and $\brho  $ is the impact parameter.
Hence
\be
\dot{r} = v_0 \sqrt{1 - r_0/r - \brho  ^2/r^2}. 
\ee

Using (\ref{intclass}) and the above definitions, we have
\be
\Sigma = \frac{32\pi G}{15 c^5} 
e_1^2 e_2^2  v_0 
\int_{r_{\rm min}}^\infty \!\!\!\! \d r \!
 \int_{0}^{b_1} \!\! \d\brho   
\frac{ (1 - r_0/r) + 11 \brho  ^2 / r^2 }
{ r^4 \sqrt{ (1- r_0/r) - \brho  ^2/r^2 } }  \, \brho
     \nonumber \\
\label{Sigclassical}
\ee
where $b_{1}=\sqrt{r^2 - r r_0}$.
Taking the integration with respect to $\brho$ first, we have that,
for constants $a,B,C,d$,
\be
\int \! \frac{C + d \brho  ^2}{ \sqrt{a - B \brho  ^2} } \brho   \d\brho  
= - \frac{ \sqrt{a - B \brho  ^2}} {B}
\left[C + \frac{2 ad}{3B} + \frac{d \brho^2}{3} \right].   \label{intsqrt}
\ee
Therefore 
\be
\Sigma = \frac{64\pi G}{9 c^5} \frac{(e_1e_2)^2  v_0}{|r_0|}  \chi  
\ee
where
\be
\chi &=& \frac{5|r_0|}{2} \int_{r_{\rm min}}^\infty
\frac{1}{r^2}
  \left(1 - \frac{r_0}{r} \right)^{3/2} \, \d r   \\
  & = &  
\frac{5}{2} \int_{x_{\rm min}}^\infty
\frac{1}{x^2} \left(1 \pm \frac{1}{x} \right)^{3/2} \, \d x
\label{Iint}
\ee
where the plus(minus) sign corresponds to an attractive(repulsive) potential.
The lower limit on the integral with respect to $r$ is the smallest $r$ 
attained in the
motion. This is zero for an attractive collision and $r_0$ for a repulsive one.
It follows that $x_{\rm min} = 0$ for an attractive collision and
$x_{\rm min} = 1$ for a repulsive one.
Consequently $\chi$
diverges for an attractive collision and one obtains $\chi = 1$ for
a repulsive collision. Hence the classical calculation (with no adjustment for quantum effects)
yields a divergent result for an attractive collision (owing to infinite acceleration in a head-on collision), and for a
repulsive collision yields
\be
\Sigma_{\rm r} = \frac{32\pi G}{9 c^5} Z_1 Z_2 e^2  m v^3   \label{sigCCclass}
\ee
where we now use $v$ to indicate $v_0$ which makes a comparison
with other results more transparent.
This is the equation first obtained by Carmeli (\cite{Carmeli1967}, eqn (4.4)). When substituted into
(\ref{PSigma}) it confirms our rough estimate~(\ref{PeeEst}).

\subsection{Quantum treatment}

We now review results of quantum scattering theory for this problem,
obtained by previous authors.  
Both Barker {\em et al.} and GG treat the Born approximation and 
give some higher-order results. We shall present 
the
Born approximation and some further observations by GG.

\begin{figure}
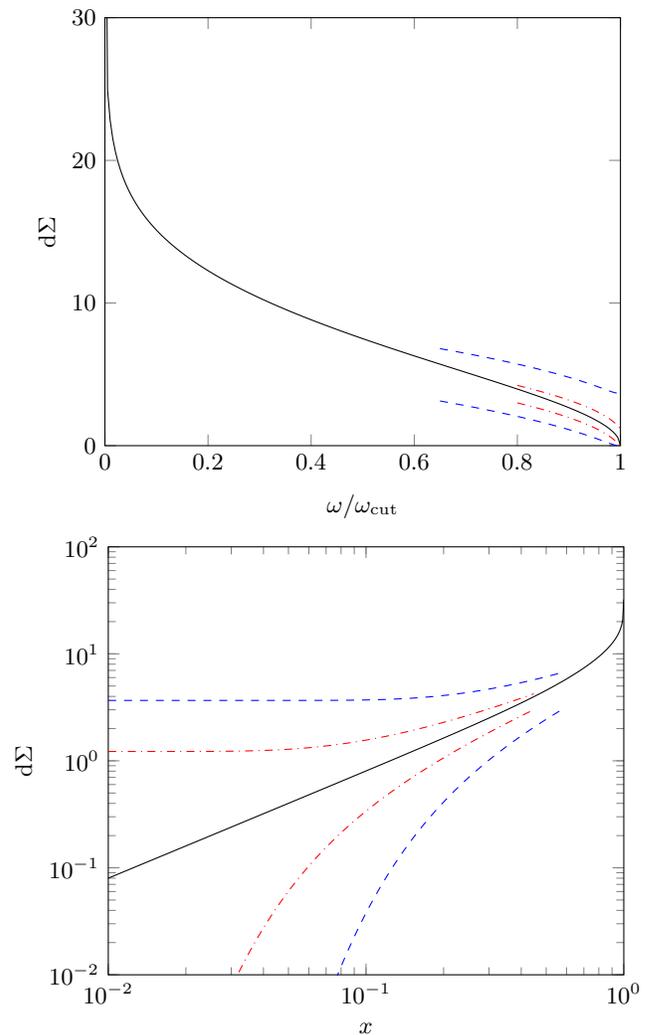

\mytikz{1}{spectrum} 
\mytikz{1}{spectrum_log} 
\caption{Spectrum of GW emission in a Coulomb collision in the first
order Born approximation for the collision 
($n^{}_{\rm B} \ll 1$), 
as given by (\ref{spectrum}). The dashed lines show the corrected spectrum near
the hard end, eqn (\protect\ref{spectrum.correction}). 
Blue dashed: $v=0.1 c$, red dash-dot: $v=0.3c$.}
\label{f.spectrum}
\end{figure}

Eqn (8) of GG is the same as eqn (10) of Barker {\em et al.}
after the replacement $(G M m/\hbar c) \rightarrow (e^2/\hbar c)$.
(This replacement is the one Barker {\em et al.} point
out after their eqn (15), except that they adopt rationalised electromagnetic units.) In our units, Barker {\em et al.}, and also GG, find
that the contribution to $\Sigma$ of the graviton frequency range
$\d\omega$, in the case of Coulomb scattering, is:
\be
{\d}\Sigma = \frac{64 G \hbar}{15 c^3 } \!\left(\frac{e_1e_2}{\hbar c}\right)^{\!2} \! 
\left[5 x + \frac{3}{2}(1+x^2) \ln \frac{1 + x}{1-x} \right]\! \hbar \d\omega
\label{spectrum}
\ee
where $x = p'/p$ is the ratio of final to initial momentum of a
particle scattering off a fixed potential. For single graviton emission
(i.e. Born approximation) we have, by conservation of energy,
$\hbar \omega = (p^2 - p'^2)/2 m = (1-x^2)p^2/2m$ so
$\hbar \d\omega = -x p^2/m$. When $\omega$ ranges from $0$ to
the hard cut-off, $x$ ranges from $1$ to $0$, so
\be
\Sigma &=& \!\frac{64 G }{15 c^5 \hbar } (e_1e_2)^2 \frac{p^2}{m}
\int_0^1 \! 5 x^2 + \frac{3}{2}x(1+x^2) \ln \frac{1 + x}{1-x}  \d x
\nonumber \\
&=& 
(160 G/9 \hbar c^5) (e_1e_2)^2 m v^2 .  \label{CBorn}
\ee
However one should keep in mind that the Born approximation is only
valid when $\nB \ll 1$ for both the initial and final momenta. At
the hard end of the spectrum $p' \rightarrow 0$ so $\nB \rightarrow \infty$.
Therefore the above formula has to be corrected at the hard end. 
This is the region where $x \rightarrow 0$. GG obtain
\be
\d\Sigma \rightarrow \pm \frac{1024 \pi G}{15 c^5} (e_1e_2)^2 \frac{ \tilde{\alpha} c}{v}
\frac{ \d\omega}
{  ( e^{\pm 2 \pi \tilde{\alpha} c/xv } - 1 ) }
\label{spectrum.correction}
\ee
where the $+$ sign is for repulsion and the $-$ sign is for attraction,
and $\tilde{\alpha} \equiv Z_1 Z_2 \alpha$.
Since $xv$ is the final speed, the corrected formula should match
the uncorrected one when the final Born parameter $\alpha c/xv \ll 1$, as indeed it does. But at the hard end, $x \rightarrow 0$, the spectrum
is different in the two cases:
\be
\d\Sigma \rightarrow \frac{1024\pi G}{15 c^5}(e_1e_2)^2 \frac{\tilde{\alpha} c}{v}
 \d\omega 
\left\{ \begin{array}{cl}
e^{-2\pi \tilde{\alpha} c/xv} & \mbox{repulsion} \\
1 & \mbox{attraction}
\end{array} \right.                     \label{dSigra}
\ee
It follows that (\ref{CBorn}) overestimates the power in the
repulsive case, and underestimates it in the attractive case;
c.f. Fig. \ref{f.spectrum}. Note also that $\d\Sigma$ scales as $(Z_1 Z_2)^3$.

The above Born approximation results apply when $\nB \ll 1$. 
Closed formulae are also available in the other limit, $\nB \gg 1$.
For repulsion one then regains the classical result
(\ref{sigCCclass}). For attraction the classical result (with no
angular momentum cut-off) diverges; the quantum 
treatment derived by GG (their eqn (17)) gives
\be
\Sigma_{\rm a} = \frac{8 G}{5 c^5} 12^{1/3} \Gamma^2({2}/{3}) \,
 Z_1 Z_2 e^2 m v^{4/3} (\tilde{\alpha} c)^{5/3}.   \label{GGattract}
\ee
where the subscript `a' stands for `attractive'.

In order to compare the various results, let us define in each case
\be
\chi \equiv \Sigma / \Sigma_{\rm r}
\ee
where $\Sigma_{\rm r}$ is given by (\ref{sigCCclass}).
From (\ref{CBorn}) one obtains
\be
\chi^{}_{\rm B} \equiv \frac{\Sigma}{\Sigma_{\rm r}} =
\frac{9 }{2\pi} \nB .       \label{cBornr}
\ee
Thus quantum effects here act to suppress the power
by a factor $9\nB / 2\pi$ compared to what would be expected classically.

Comparing now attraction and repulsion in the low-velocity limit, we have
\be
\chi_{\rm a} \equiv \frac{\Sigma_{\rm a}}{\Sigma_{\rm r}}
\simeq 0.6013 ( \tilde{\alpha} c/v)^{5/3} = 0.6013 \, \nB^{5/3}.
\label{cBorna}
\ee
The power in the attractive case greatly exceeds that in the repulsive
case for low $v$. 
This is because the relevant speed
for the attractive case is not the incident speed but the speed
at closest approach. For a classical trajectory
at angular momentum $L$, the speed at closest approach is
approximately $\nB v \hbar / L = \tilde{\alpha} c \hbar / L$
in the limit $\nB \gg L/\hbar$.
The scaling $v^{4/3}$ exhibited in (\ref{GGattract}) can be
interpreted as the cube
of a velocity which makes a compromise (roughly a geometric mean)
between $v$ and $\nB v$. 

\begin{figure}
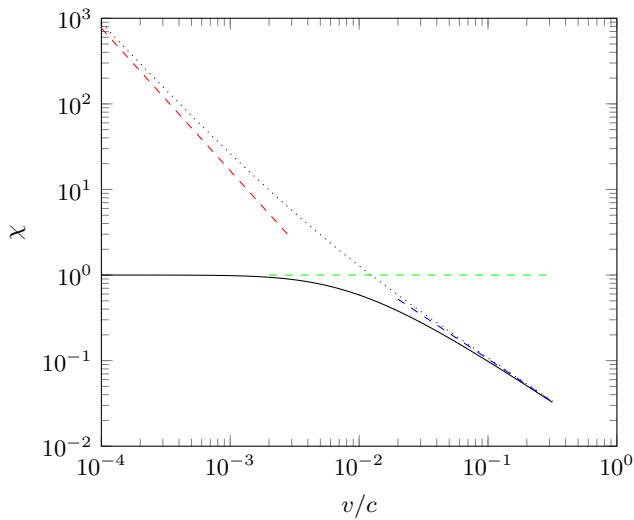

\mytikz{1}{chiCoul}
\caption{Predictions for GW radiation in Coulomb collisions.
The dashed lines show the limiting cases as described by 
(\ref{sigCCclass}) and (\ref{cBornr}) (low $v$) and (\ref{cBorna}) (high $v$).
The full (dotted) line shows the predictions of the modified classical
method described in section \ref{s.semi}
(eqns (\ref{chirlam}), (\ref{chialam})).
The horizontal axis is $\tilde{\alpha}/{\protect\nB}$; 
this is equal to $v/c$ in the case of electron collisions.}
\label{f.chiCoul}
\end{figure}

The predictions of (\ref{sigCCclass}), (\ref{cBornr}) and (\ref{cBorna}) 
are plotted as dashed lines on figure \ref{f.chiCoul}.

\subsection{Soft photon theorem}  \label{s.softC}

The soft photon theorem has to be applied with caution in the case of Coulomb collisions owing to the divergence of the collision cross-section term
in (\ref{Sigsoft}). 
That is, the quantity
\be
\tilde{\sigma} \equiv \int \dby{\sigma}{\Omega} \sin^2 \theta \d\Omega
\ee
diverges. Therefore the approximations invoked in the theorem do not hold
in the case of the Coulomb potential. The problem is the long-range
nature of $1/r$; similar difficulties arising in other scattering
problems associated with this potential. In practice in a plasma 
there will be Debye screening which leads to a Yukawa potential. One then
finds that $\tilde{\sigma} \sim v^{-4} \ln v$ in the limit of small
screening.

The soft photon/graviton theorem does not give the whole emitted power
and one only expects order-of-magnitude agreement
with the full $\Sigma$ in general. However by judicious choice of the
cut-off $\Lambda$ one may expect to reproduce the full $\Sigma$ to within a
factor 2 for the repulsive case.
For Coulomb collisions there are two relevant frequency scales: the
inverse of the collision time $|r_0|/v$ (where $v$ is the maximum speed), 
and the
hard cut-off at $K/\hbar$ where $K = (1/2) mv^2$.
If we use as $\Lambda$ the smaller of $\hbar v_0/r_0$ and $K$, and take 
\be
\tilde{\sigma} \simeq 32 \pi \alpha^2 \left( \hbar / \mu c \right)^2 (c/v)^4,
\ee
then the behaviour shown in figure \ref{f.chiCoul} for repulsive collisions 
is reproduced 
by the formula (\ref{Sigsoft}) in both limits of low and high $v_0$. 

For attractive collisions the soft photon theorem is less successful, but gives
a good estimate at high $v_0$ (low Born parameter).

\subsection{Modified classical}   \label{s.semi}

As noted in section \ref{s.not}, our modified classical method
of calculation is an adjustment of the classical integrals
so as to yield a reasonable approximation. 
We consider the effect of adjusting $x_{\rm min}$ in (\ref{Iint}), giving
\be
\chi_r(\lambda) &\! = \!&
\frac{5}{2} \int_{1 + \lambda}^\infty \! \frac{\d x}{x^2}
 \left[1 - \frac{1}{x} \right]^{3/2} \!
= 1 - \left[1 + \frac{1}{\lambda} \right]^{-5/2}    \;\;   \label{chirlam}   \\
\chi_a(\lambda) &\!=\!&
\frac{5}{2} \int_{\lambda}^\infty \! \frac{\d x}{x^2}
 \left[1 + \frac{1}{x} \right]^{3/2} \!
= -1 + \left[1 + \frac{1}{\lambda} \right]^{5/2}    \;\;    \label{chialam}
\ee
where $\lambda$ is a parameter which one would expect to 
be of the order of the de Broglie wavelength divided by $|r_0|$.

Defining $\lambda_{\rm dB} \equiv 2 \pi \hbar / \mu v_0$, one finds
$
 \lambda_{\rm dB} / |r_0| = \pi / \nB.
$
By setting the parameter value
\be
 \lambda = 0.5515 \pi / \nB    \label{lambdar}
\ee 
one finds that (\ref{chirlam})
reproduces the known results in both classical and quantum limits,
and gives reasonable behaviour at all $v < c$, see Fig. \ref{f.chiCoul}.

For attractive collisions the distance scale where quantum effects must
be allowed-for is not simply $|r_0|$ but may be considerably smaller. 
By solving for $r$ the equation $r = h/\mu v$ with 
$v = v_0 (1 + |r_0|/r)^{1/2}$ one finds $r \simeq \pi \lambda_{\rm C}/\alpha$
where $\lambda_{\rm C} = h / \mu c$ is the Compton wavelength. We mention
this merely to indicate that the attractive case is less straightforward.
We shall choose the parameter $\lambda$ so as to match 
(\ref{cBorna}) in the low-velocity
limit and (\ref{cBornr}) in the high-velocity limit. 
We also have a further piece of information provided by (\ref{dSigra}),
namely that $\chi$ approaches
the asymptote from above at small Born parameter in the attractive case.
These constraints are achieved by adopting (for example)
\be 
\lambda = \left( 5.20 + 1.84 \, \nB \right)^{1/3} / \nB.    \label{lambdaa}
\ee
The result is shown in Fig. \ref{f.chiCoul}.


Eqns (\ref{chirlam})--(\ref{lambdaa}) together
provide a formula for $\chi$  which is approximately
valid at all collision speeds $v$. This $\chi$ is the ``Gaunt factor" 
for the total (i.e. integrated over frequency) emission. It allows one
to obtain $\Sigma$ by taking $\Sigma_{\rm r}$ given by (\ref{sigCCclass}) and multiplying by a correction factor.

The task of calculating gravitational scattering amplitudes and other
observables in full is presented in Kosower {\em at al.}.\cite{Kosower2019}
They offer at once an extensive review, a tutorial, and original
contributions. There is a procedure to systematize the approximations which
allows the classical limit to be taken. Such a calculation remains lengthy.
The crude-but-reasonably-accurate method presented
here serves partly as a simple formula to apply to study of plasmas and clusters (c.f. section \ref{s.sun}), and partly to facilitate the checking of more advanced methods.

\section{Power and energy for a given scattering event} \label{s.single}

So far we have not treated the motion during individual scattering events,
because it was convenient to integrate over impact parameter. We now treat
individual events of given $b,\,v_0$. 
We shall present the gravitational (Keplerian), i.e. attractive case.

The orbit can be described by the parameters $b,\,v_0$ or by a number of
other pairs, including $E,L$ (energy and angular momentum, both conserved) and
$a,e$ where $a \equiv GM/v_0^2 = - r_0/2$ and
$e$ is the eccentricity defined by
\be
 e = \sqrt{1 + b^2/a^2} \,.
\ee
For a hyperbolic orbit one then finds that the distance of closest approach is
\be
r_{\rm min}  = a(e-1) = b \sqrt{ \frac{e-1}{e+1} }  \label{rmine}
\ee
and 
$
e = -1 / \cos \phi_0
$
where $\phi_0$ is half the total change in azimuthal angle during the encounter
(the deflection angle is $2 \phi_0 - \pi$). 

On a classical model under the adopted assumptions (i.e. motion in a
$1/r$ potential), the GW power during the scattering process is given 
by (\ref{Powervv}), which, after using the conservation laws (\ref{conserve}), gives an expression in terms of $r$ and constants.
Turner gives the following formula (eqn (24) of \cite{Turner1977}):	
\be
P &\!\!=\!& \frac{8G^4}{15 c^5}\frac{M^3 \mu^2 (1 + e \cos \phi)^4}{[(1+e)r_{\rm min}]^5}
[e^2 \sin^2 \! \phi + 12(1 + e \cos\phi)^2]  \nonumber \\
\label{PowerT}
\ee
where $\phi$ is the azimuthal angle taken from $\phi=0$ at periastron (the point where $r = r_{\rm min}$). Thus $\phi$ goes from $-\phi_0$ initially to
$\phi_0$ finally.

Capozziello {\em et al.} give (eqn (21) of \cite{Capozziello2008}):
\be
P = \frac{32 G L^6 \mu^2}{45 c^5 b^8}f(\phi_0,\psi)
\label{PowerC}
\ee
where
\be
f(\phi_0,\psi) = \frac{
\sin^4\left( \phi_0 \!-\! \psi/2 \right) 
\sin^4\!\left( \psi / 2 \right) }
{\tan^2 \phi_0 \sin^6 \phi_0} 
\left[
	150 + 72 \cos2\phi_0 \right. \nonumber\\
\left. 
+ 66 \cos 2 (\phi_0 \!-\! \psi) - 144 (\cos (2\phi_0 \!-\! \psi) + \cos \psi)
	\right]. \;\;\;\;
\ee
(This formula is quoted incorrectly in \cite{Vittori2012} where there is a sign error in
the last term). Here $\psi \equiv \phi + \phi_0$ (thus $\psi$ goes from 0 initially to $2 \phi_0$ finally). 
If we express $f$ in terms of $\phi$ rather than $\psi$, it simplifies a little:
\be
f = 
\frac{3}{8} \frac{\left(\cos {\phi_0}-\cos {\phi}\right)^4 }
{\tan^2 \phi_0 \sin^6 \phi_0} 
\left[
25
+12 \cos 2 {\phi_0}  \right. \nonumber\\ \left.
 - 48 \cos{\phi_0} \cos {\phi}
+11 \cos 2{\phi}
\right] .
\ee
Equations (\ref{Powervv}), (\ref{PowerT}) and (\ref{PowerC}) give three ways of expressing the same result.
They are all equivalent, which one may confirm by employing
$
r = {b \sin \phi_0} / ({\cos \phi - \cos \phi_0}) 
$
(a standard result of orbital mechanics). 
	
The integral of $P$ over time is conveniently done by converting to an integral over $\phi$. The result was first obtained by Turner:
\be
\Delta E = \frac{8 G^{7/2}}{15 c^5} \frac{M^{1/2} m_1^2 m_2^2}{r_{\rm min}^{7/2}} g(e)            \label{DETurner}
\ee
with
\be
g(e) = \frac{ \phi_0 (24 + 73 e^2 + \frac{37}{4} e^4) + 
\frac{\sqrt{e^2-1\,}}{12} (602 +673 e^2)}
{ (1+e)^{7/2} }            \label{geTurner}
\ee
(correcting an earlier calculation of Hansen). 
In order to bring out the comparison with (\ref{sigCCclass}), note that
\be
 \frac{8 G^{7/2}}{15 c^5} \frac{M^{1/2} m_1^2 m_2^2}{((e+1)r_{\rm min})^{7/2}}
= \frac{8  G}{15 c^5} \frac{G M \mu^2  v_0^3}{ b^2 (e^2 - 1)^{5/2}} .
\ee

Dehnen and Ghaboussi'e result (eqn (7) of \cite{Dehnen1985}) is
\be
\Delta E = 
\frac{ 8 G (e_1 e_2)^2}{15 c^5}
\frac{\mu E^2}{L^3} 
\left[	
(37 + 366 z^2 + 425 z^4) \phi_0 + \right. \nonumber\\
\left. (673/3 + 425 z^2) z \right]
\ee
where
$
z \equiv -\!\cot \phi_0 \!=\! (e^2 - 1)^{-1/2} .
$
This agrees with Turner after one makes the substitution $e_1 e_2 \rightarrow - G m_1 m_2$.

The total scattered energy was also obtained by Capozziello {\em et al.} Their expression
is consistent with Turner's if one handles the term $\sqrt{e^2-1}$ correctly.
It must be taken positive, which means it is equal to $-\tan \phi_0$ not $\tan\phi_0$
when $e > 1$. Also, \cite{Vittori2012} give a result a factor 4 larger than that of \cite{Capozziello2008}. In view of
these issues a further check is useful. We completed the calculation independently
and agree with Turner (and therefore also Dehnen and Ghaboussi)
and with Capozziello {\em et al.} as long as the correct
sign is taken, as just noted. 

\section{Classical collisions with angular momentum cut-off} \label{s.cutoff}

So far we have surveyed or confirmed existing work, and contributed a small
extension in the modified classical method. The remainder
of our discussion is mostly new. 

Rather than taking the integral (\ref{intclass}) over all impact parameters,
we now place a lower limit on $b$. This will be useful for two purposes.
First, the influence of quantum mechanics on collision cross-sections can sometimes be
estimated by imposing a low angular momentum cut-off, at a value of order
$\hbar$, on a classical collision integral. Secondly, for
attractive collisions the low angular momentum limit has to be considered
separately in any case. This is 
because the approximation that the orbit is almost unaffected by the 
radiation breaks down.

\begin{figure}
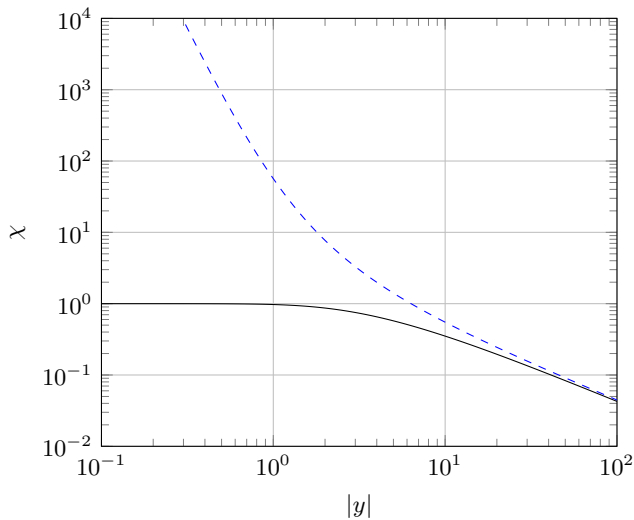

	\mytikz{1}{chiL}
	\caption{$\chi(L,v_0)$ given by (\ref{chiL}) for attractive (upper line, dashed)
		and repulsive (lower line, full) collisions.}
	\label{f.chiL}
\end{figure}

In place of eqn (\ref{intclass}) we introduce
\be
\Sigma(L,v_0)
\equiv 2  \int_{r_{\rm min}}^\infty \frac{\d r}{|\dot{r}|} 
\int_{L/mv}^{\brho_{1}} 2 \pi \brho \,  \d\brho   \, L_{\rm GW}
\label{intLclass}  
\ee
where $L$ is the cut-off and the notation on the left hand side is to indicate explicitly that the
result is a function of the cut-off angular momentum $L$ as well as $v_0$.
Then in (\ref{Sigclassical}) we replace the lower limit $0$ on the $b$ integral
by $\brho_0 = L / mv_0$, and $r_{\rm min}$ is given by (\ref{rmine}) (and by
\ref{rminL}).
After using (\ref{intsqrt}) we obtain
\be
\Sigma(L,v_0) = \frac{64\pi G}{9 c^5} \frac{(e_1e_2)^2  v_0}{|r_0|}  \chi(L,v_0)  \label{ILdef}
\ee
where
\be
\chi = \frac{|r_0|}{10} \int_{r_{\rm min}}^\infty  \!
\frac{\d r}{r^2}
\left[ 25 \left(1 \!-\! \frac{r_0}{r}\right) + 11 \frac{\brho_0^2}{r^2} \right]
\left[1 - \frac{r_0}{r} - \frac{\brho  _0^2}{r^2} \right]^{1/2}   \nonumber
\ee
The lower limit on this integral is the smallest $r$ attained in the
motion when the impact parameter is $\brho  _0$. This is
\be
r_{\rm min} = \frac{1}{2}\left(r_0 + \sqrt{r_0^2 + 4 \brho  _0^2}\right)   \label{rminL}
\ee
where the positive square root should be taken.
(For $L = 0$ this gives $r_{\rm min} = r_0$ for a repulsive collision and
$r_{\rm min}=0$ for an attractive collision.) One finds
\be
\chi(L,v_0) &=& \frac{1}{80 |y^5|}
\left[ 6(1+y^2)(85+37 y^2) \left(\frac{\pi}{2} - \cot^{-1} \! y \right)
\right. \nonumber\\ 
&& \left. - 510y - 562 y^3 \rule{0ex}{2.2ex}\right] 
\label{chiL}
\ee
where 
\be
y \equiv \frac{L v_0}{e_1 e_2} = \pm \sqrt{e^2-1},
\qquad |y|  = \frac{ L/\hbar }{ \nB } 
\ee 
where the negative square root is taken for the attractive case.
$\chi(L,v_0)$ is plotted as a function of $y$ in figure \ref{f.chiL}.
It is remarkable that this $\chi$ is a function of eccentricity alone.

One finds
\be
\chi(L, v_0) \rightarrow \left\{ \begin{array}{ccl}
	1 && y \ll 1, \; y > 0 \\
	(51 \pi / 8) y^{-5}   && |y| \ll 1, y < 0 \\
	(111 \pi / 80) y^{-1} && |y| \gg 1
\end{array} \right.
\ee
Positive $y$ means the potential is repulsive. 
At small $y$ the result is then independent
of $L$ and reproduces the classical calculation without any angular momentum
cut-off. This is because at small initial velocities the particles do not approach closely in a repulsive potential. At large $y$ the result exactly
reproduces the first
order Born approximation (\ref{CBorn}) in the limit if we take
\be
L = \frac{37 \pi^2}{120}  \hbar \simeq 3.043 \, \hbar .
\ee
It follows that 
$\Sigma(3.04 \hbar, v_0)$ can be taken as a reasonable approximation to the exact result
(i.e. a quantum scattering calculation to all orders) for GW scattering during Coulomb collisions on a repulsive potential, for any collision energy in the non-relativistic limit.
In other words,
{\em for repulsive Coulomb collisions the complete quantum scattering prediction (summed over all orders or Feynman diagrams)
closely matches a classical prediction
in which low angular momentum states do not contribute at all}. The phrase
`closely matches' here signifies exact agreement in the limits of large
or small $\nB$, and agreement at some
unknown accuracy of order 10\% in the case $\nB \sim 1$.

For an attractive potential 
$\Sigma(3.04 \hbar,v_0)$ produces the correct cross-section
at high $|y|$ but not at low $|y|$. In other words, for an attractive 
Coulomb collision
it is not sufficient merely to place a lower bound on the angular momentum in order
to approximate the quantum physics of a collision at low energy.

\section{Gravitating clusters}  \label{s.cluster}

In section \ref{s.total} 
we discussed the total emission cross section,
integrating over all impact parameters. For emission from a plasma this
is a useful quantity, but for gravitational scattering in general
it is not because 
the approximations break down at low $L$ in an attractive potential. 
Various situations can arise.
Astrophysical bodies are generally
not point-like and can crash into each other or otherwise merge.
Also, even on a point-like model there can be radiative capture.
This happens when
\be
\Delta E > \frac{1}{2} \mu v_0^2.    \label{DEE}
\ee
That is, the emitted energy is larger than the initial energy in the binary system, with the result that an initially unbound pair finishes in a bound state. In a bound state the pair subsequently follows an almost periodic, almost elliptical orbit, gradually losing more energy until the bodies coalesce.

In order to treat a gravitating cluster,
we separate the scattering events into those where the
bodies emerge to infinity, and those where there is gravitational capture
owing to the gravitational radiation. We will employ the condition
(\ref{DEE}) to separate the two cases, which is valid at a low density of
pairs but not at higher density where three-body effects tend to reduce the 
capture rate. \cite{Vaskonen2020}

Using (\ref{DETurner}) on the left hand side of (\ref{DEE}) we find that
the limiting case (where $\Delta E = E$) is given by 
\be
e - 1 = \left( \frac{16}{15} \frac{\mu}{M} \frac{v_0^5}{c^5} g(e) \right)^{2/7} .
\label{egminus}
\ee
This method of calculation is approximate since for such collisions the 
outgoing value of $e$ will not be equal to the initial value, but it gives
a reasonable estimate. Eqn (\ref{egminus}) has $g(e)$ on the right hand side
so it is an implicit equation for $e$ with no analytical solution. But
we observe that for $v_0 \ll c$ one has $e - 1 \ll 1$ as one would expect:
$e = 1$ is the parabolic orbit where $E = 0$. In this case we can use 
$g(1)$ on the right hand side, obtaining
\be
e - 1 \simeq \left( \frac{85 \pi }{6 \sqrt{2}} \frac{\mu}{M} \frac{v_0^5}{c^5}  \right)^{2/7} .         \label{eminus}
\ee
This agrees with eqn (17) of \cite{OLeary2009}.
Non-captured orbits have $e-1$ larger than this. We should now note two
consistency checks. For the Newtonian potential to be valid we
require $r_{\rm min} \gg R_{\rm S} = 2 GM/c^2$ (the Schwarzschild radius). This
yields the condition
\be
e - 1 \gg 2 v_0^2 / c^2.
\ee
This is comfortably satisfied by (\ref{eminus}) for $v_0 \ll c$. Also
for non-relativistic mechanics we require $v_{\rm max} \ll c$. Conservation
of angular momentum gives $r_{\rm min} v_{\rm max} = b v_0$ and one obtains
\be
\frac{e-1}{e+1} \gg \frac{v_0^2}{c^2}.        \label{emep}
\ee
Since $e+1 > 2$ this is a stronger condition than the previous
one, but still comfortably satisfied.

We have in (\ref{eminus}) an expression for the minimum 
eccentricity, at any given $v_0$, for non-captured orbits. Since $e -1 \ll 1$
we can use $y \equiv -\sqrt{e^2-1} \simeq -\sqrt{2} (e-1)^{1/2}$, and since
this is small we can use the small $|y|$ limit of 
eqn (\ref{chiL}), giving
\be
\chi(y) \simeq \frac{51 \pi}{32 \sqrt{2}}
 \left( \frac{6 \sqrt{2}}{85 \pi } \frac{M}{\mu} \frac{c^5}{v_0^5} \right)^{5/7}.
\ee
Hence the total cross-section for emission of gravitational wave energy during
hyperbolic (i.e. non-captured) encounters, in a low-density, low-velocity
gravitating cluster is 
\be
\Sigma = \frac{\pi }{5} \left( \frac{340 \pi}{3 \sqrt{2}} \right)^{2/7}
	\frac{G M}{c^2} G m_1 m_2 \left( \frac{\mu}{M}\right)^{2/7}
	\left( \frac{c}{v} \right)^{4/7}  \, .
\label{SigCHE}
\ee

As an example, consider information furnished by O'Leary {\em et al.}.
They remark, ``20,000 BHs are expected to have segregated into the 
inner $\sim$1 pc of the Milky Way".\cite{OLeary2009} The number density distributions
in their figure 1 give $n \simeq n_0 (r_0/r)^2$ for $r_0 < r < 0.3\,$pc,
where $r$ is the distance from the centre of the galaxy, $n_0 \simeq
10^{10}\,{\rm pc}^{-3}$ and $r_0 = 3 \times 10^{-4}\,$pc. 
They propose black holes in the mass range 5 to 15
$M_\odot$ and encounters with initial
relative speeds of order $v \sim 1000\,$km/s. Putting these values
into (\ref{SigCHE}) and (\ref{PSigma}) we obtain a total power from
close hyperbolic encounters of black holes in the galactic centre
of order $10^{25}\,$watt after averaging over times long enough for
many encounters. 

\section{The gravitational radiation of the Sun}  \label{s.sun}

Consider now a plasma in thermal equilibrium at the density and temperature of
the core of the Sun---c.f. table \ref{t.sun}. The thermal energy $\kB T_{\rm 
	core} \simeq 1.35\,$keV is about twice the Fermi energy of the electrons, and 
therefore the electron gas is non-degenerate to reasonable approximation. 
Each electron or proton has a kinetic energy of order
$\kB T$ and the r.m.s. energy is approximately $E_Q \simeq 2\kB T$.

\begin{table}
	\begin{tabular}{llll}
		$T_{\rm core}$ & \multicolumn{2}{c}{$1.57 \times 10^7$ K} \\
		$(3/2) \kB T_{\rm core}$ & \multicolumn{2}{c}{$2.03$ keV} \\
		Coulomb distance $r_0$ & \multicolumn{2}{c}{$0.7$ pm} \\
		plasma wavelength $\lambdabar$ & \multicolumn{2}{c}{$640\,$pm}\\
		Debye (screening) length $\lambda_D$ & \multicolumn{2}{c}{$23\,$pm}\\
		& electrons & protons \\
		mean separation  &  25 pm & 32 pm \\
		$\lambda_{\rm th} = \hbar \sqrt{2\pi/m \kB T}$ &  18.8 pm &  $0.43$ pm \\
		$\lambdabar_{\rm dB} = \hbar / \sqrt{2 m E}$  &  4.3 pm &   $0.10$ pm
	\end{tabular}
	\caption{Some properties of the solar core. pm = picometre.
		$\lambda_{\rm th}$ is defined such that $n \lambda_{\rm th}^3$ is the onset of degeneracy.
		$\lambdabar_{\rm dB}$ is the distance over which a de Broglie wave acquires a phase of one radian,
		for a particle of energy $E = (3/2) \kB T_{\rm core}$.}
	\label{t.sun}
\end{table}

Gravitational bremsstrahlung in the Sun arises mainly from
collisions among electrons, protons and $^4$He nuclei. 
We shall present the result of integrating the emission over the Sun, 
treating the collisions as Coulomb collisions. This ignores the
effect of Debye screening and therefore cannot be taken as an accurate
value for the actual situation. But the Debye screening is not
expected to change the overall result by as much as an order of magnitude.
Therefore
a calculation using the unscreened potential is a useful indicator,
and also serves to establish which regime of behaviour (low or high
Born parameter, attractive or repulsive collisions) dominates.

In the solar core we have $|\nB| \simeq 0.06$ for collisions involving
electrons. It was remarked
by GG that the emission is therefore substantially reduced
below the value predicted by the classical calculation (\ref{sigCCclass}) 
(we find one order of magnitude below, not two as suggested by GG). 
We observe also that it is important to include
the attractive (ep and eHe) collisions as well as the repulsive ones.

The total power is obtained by adding the contributions from the various
types of collision, integrated over the temperature and density
distribution of the Sun. In order to perform such an integral, we
adopted the distributions given by the Standard Solar Model.\cite{TurckChieze2016,Vinyoles2017}
The result of the numerical integration is indicated in table
\ref{t.power}. We find that the total power is 76 MW (in the absence
of Debye screening). 
This is the first time this power has been calculated with better than
order of magnitude accuracy. (The previous best estimate was that of
GG who estimated the order of magnitude as 10 MW). It follows that
the GW power of the Sun is $76 \pm 20\,$MW, where the uncertainty 
is mostly owing to the as-yet-uncalculated impact of Debye screening.

\begin{table}
\begin{tabular}{l|ccc}
& e & p & He$^{++}$ \\
\hline
e & 26  \\
p & 29 & 0.096 \\
He$^{++}$ & 21 & 0.048 & 0.004 \\
\hline
total & 76 & 0.14 & 0.004 \\
\end{tabular}
\caption{Total GW power, in megawatts (MW), from the main types of Coulomb collision
	in the Sun.}
\label{t.power}
\end{table}

It is noteworthy that ee, ep and eHe collisions make almost equal
contributions. If it were not for the quantum effects, it would not
be so. For if we simply set $\chi = 1$ for all the processes, then
one finds the ee collisions dominate owing to their smaller reduced
mass, leading to higher velocities. The value $\chi=1$ also leads
to a total power 10 times larger, indicating that
the quantum effects are important for the
conditions of the Sun.
Note also that the increased emission for attractive, as compared
with repulsive, collisions also raises the contribution of ep
and eHe collisions a little, compared with ee.

\ifodd 0
\begin{figure}
\mytikzoct{0.4}{nTrfig.tikz}
\caption{The quantity $n_e^2 T^2 r^2$ (appearing in the integrand when
eqn (\ref{GG}) is integrated over volume) as a function of radius in the Sun,
in arbitrary units.}
\label{f.TTnn}
\end{figure}
\fi

From the above one may deduce that there is gravitational noise in
the Sun with an rms strain amplitude of order $10^{-41}$
at $10^{18}\,$Hz owing to Coulomb collisions. This is the dominant
source of gravitational noise in the solar system 
at this frequency.
The energy density of this radiation arriving at Earth is 
of order $10^{-24}\,$Wm$^{-3}$. This is similar to the energy density
of relic gravitational waves in the frequency band up to GHz 
thought to be present owing to early-universe processes.\cite{09LIGO,06Coward,Steane2017}
Owing to their lower frequency, the latter will have larger observable effects.

\section{Conclusion}  \label{s.conc}

In conclusion, we have achieved the five aims set out at the end of section
\ref{s.history}. We have
reviewed studies of gravitational bremsstrahlung during
Coulomb collisions and presented a formula, based on
semi-classical physical reasoning, which is able
to reproduce, approximately, the predictions of a full (i.e. quantum) treatment of the total emitted power at any value of the Born parameter, in the
non-relativistic limit.
Equations  (\ref{chirlam})--(\ref{lambdaa}) 
allow one to calculate
the energy cross-section with high accuracy in certain
limits and with $\sim\!10$\% accuracy in general. One can
thus obtain the power averaged over many collisions in
a homogeneous fluid. As an example, we have applied
these equations to a treatment of the Sun, obtaining the total emitted power in the 
approximation where Debye screening is neglected. 

Eqn (\ref{chiL}) (combined with (\ref{sigCCclass}))
gives the cross-section for gravitational wave emission
in the classical (high
Born parameter) limit for collisions at a given initial velocity 
after integrating over impact parameters above a lower limit set by
a given angular momentum. This has not previously been calculated.
We have used it to obtain, in eqn (\ref{SigCHE}), the total 
cross section for emission of GW energy
during close hyperbolic encounters where capture does not
occur. This can be used to calculate, for example, the time-averaged
emission from galactic nuclei by this process.

It has recently been suggested that black hole collisions in the early universe 
made a non-negligible contribution to the stochastic gravitational background in 
the present. One may ask whether Coulomb collisions in the very early universe
made a further non-negligible contribution. I have attempted an estimate
of this (unpublished); the estimate suggests that the contribution is negligible
but it would be interesting nonetheless to look into this more fully.

\bibliographystyle{plain}
\bibliography{../gravitycosmology}

\end{document}